# A review of thermal transport and electronic properties of borophene

Dengfeng Li[*], Ying Chen, Jia He, Qiqi Tang, Chengyong Zhong and Guangqian Ding

*School of Science, Chongqing University of Posts and Telecommunications, Chongqing, 400065, China.*

*Corresponding author, E-Mail address: lidf@cqupt.edu.cn

**Abstract**

In recent years, two-dimensional boron sheets (borophene) have been experimentally synthesized and theoretically observed as promising conductor or transistor with novel thermal and electronic properties. We first give a general survey of some notable electronic properties of borophene, including the superconductivity and topological characters. We then mainly review the basic approaches, thermal transport, as well as the mechanical properties of borophene with different configurations. This review gives a general understanding of some of the crucial thermal transport and electronic properties of borophene, and also calls for further experimental investigations and applications on certain scientific community.

## 1 Introduction

The fundamental physical properties of materials, such as thermal and electronic properties, are crucial to the performance and longevity of nano-devices. As the miniaturization of electronic devices, such as transistors, the silicon-based devices are not well suitable for the high standard of electronic transport and thermal diffusion in nano-devices. However, the discovery of graphene brought some light to this puzzle for its ultra-fast electronic transport properties and high thermal conductivity[1, 2]. In past decades, the discovery of fullerene, carbon nanotube and graphene make the carbon materials are one of the most promising candidates for the future electronic devices[3-5].

Different from the huge success of carbon materials, the boron materials receive relatively less attention for the large challenge in experimental realization. Boron has three valence electrons, which makes the bonding mechanism is extremely complicated, thus, the structural diversity and physical and chemical complexity are beyond those of carbon. Bulk boron is characterized by three-dimensional polyhedral structural features, in which the dominant motif is isolated $B_{12}$ icosahedron[6]. However, the electron-deficient character of boron leads to the 2D boron structures present higher stability. Prior to the study of 2D boron sheets, the boron clusters had been found to be planar or quasi-planar[7-9], laying the foundations for the discovery of 2D boron nanostructures. It was reported that 2D boron sheets can be constructed via removing atoms in the triangular lattice or adding atoms into the hexagonal lattice[10]. A structural parameter, named hexagonal vacancy density $\eta$, is used to describe the ratio of hexagon holes to the number of atomic sites in the original triangular lattice within one unit cell. 2D boron sheets with different $\eta$ possess different electronic properties, such as high anisotropy, and superconductivity[11-13]. Besides, from first-principles calculations, there are lots of proposed 2D boron allotropes which are not belong to above structural paradigm have been investigated extensively with different properties[14-19]. The synthetic mechanism of 2D boron on metal substrates or metal boride substrates had been discussed with first-principles calculations[20,21], and inspiringly, there are few 2D boron sheets have been confirmed in experiments, such as $\beta_{12}$, $\chi_3$[22] and $\delta_6$[23].

Since the successful realization of 2D boron sheets on Ag (111) substrate, 2D boron materials have raised more and more attention[24]. For convenience, we will call 2D boron sheet as borophene, in analogue to graphene. For better understanding of borophene, one of the most considered issues is the thermal properties of the short and strong covalent bonds in boron materials. Meanwhile, considering the bonding complexity and structural diversity, the electronic properties in borophene would be a great interest. As we know, the boron atom is light atom, and the intrinsic electron-phonon coupling is strong, therefore, the superconductivity in boron materials is also a very hot topic. Additionally, the negligible spin-orbit coupling makes the topological classifications in boron materials should be different from the strong spin-orbit coupling materials as discussed in carbon materials[25, 26]. In this review, at first, we will introduce the notable electronic properties of borophene in part 2, e.g., the superconductivity and topological properties. Then, we will mainly focus on the basic theories and thermal transport properties of borophene in part 3. The last part is the conclusion. Although, some efforts were dedicated to the electronic properties of borophene, people were mainly interested in thermal transport. Hence, after a brief introduction of electronic properties, we will mainly discuss the thermal transport properties of borophene. We would refer to most of the researches and achievements made in recent years and organize in a comparative way, providing a better understanding of these materials for further studies.

## 2 Electronic properties

The nature of electron-deficient of boron atoms makes the structural complexity of its allotropes is greater than that of carbon, leading very abundant physical properties, some of them even superior to carbon materials, such as the superconductivity and topological properties.

Compared with carbon outer-shell valence electron $2s^2 2p^2$, boron lacks one electron with $2s^2 2p^1$. Therefore, the stable configuration of honeycomb lattice of carbon is unstable for boron. Moreover, this essential difference leads to completely different structural landscape for boron. Because of lacking one electron, the bonding mechanism in boron sheets is usual the mixture of two-center bonding and three-center bonding. The recent experimental realization of 2D boron sheets, e.g. triangular boron, $\beta_{12}$, and $\chi_3$, are metal. The analysis of the origin of the stability and metallic properties of boron sheet were discussed by Tang *et.al.*[10]. They divided the bonding state of the boron sheet into in-plane part (contributed by orbital s, $p_x$ and $p_y$) and out-of-plane part (contributed by orbital $p_z$). The in-plane part forms σ bonding and anti-bonding, the out-of-plane part forms π bonding and anti-bonding. In normally hexagonal lattice, the three electrons not fulfill three σ bonding states, but half-fill π bonding, which makes the hexagonal lattice of boron unstable and metallic, and acting as acceptors. However, for the flat triangular lattice has the Fermi level over the crossing point of σ bonding and antibonding states. The over-occupied σ antibonding states also make this sheet unstable and metallic, meanwhile, acting as donors. Ideally, the hexagon mixed with triangle with the highest stability should have electrons fill all available in-plane bonding states and the low-energy out-of-plane states, which will lead to a metallic system.

## 2.1 Superconductivity

According to the BCS theory, metals composed of light-weight elements are beneficial for increasing the superconducting transition temperature (Tc), because the Debye temperatures of these metals are typically high[27]. The prevailing two-dimensional materials, such as graphene, silicene and phosphorene, are not well suited to produce high-Tc for their semimetal and semiconductor electronic properties and the weak electron-phonon coupling. However, the lighter boron materials theoretically have much stronger electron-phonon coupling; meanwhile, the metallicity of two-dimensional boron is another critical advantage for studying the superconductivity, without needing of any external charge carrier doping.

The high-Tc of binary boride, such as $MgB_2$[28], and the successful experimental synthesis of two-dimensional boron further impel the interesting of finding the superconducting states in elemental boron materials. Considering the short history and insufficient experimental studies of two-dimensional boron allotropes, for now, people study the superconductivity of boron materials mainly from the theoretical perspective view[19]. Using *ab initio* evolutionary algorithm and first-principles calculations, Zhao *et al* proposed five energetically metastable 2D boron structures, and found the superconductivity are ubiquitous in these 2D boron materials[29]. As shown in Figure 1, Penev *et al*[12] and Zhao *et al*[13] discovered the superconductivity and Tc is a V-like function of the hexagonal hole density η in the triangular lattice. Recently, there are much attentions paid to the experimentally reported 2D boron structure, triangular boron, $β_{12}$ and $χ3$. Xiao *et al* found the superconductivity in triangular 2D boron can be significantly enhanced by strain and carrier doping[30]. Gao *et al* found the electron-phonon coupling constants in the $β_{12}$ and $χ_3$ are larger than that in $MgB_2$; the Tc were determined to be 18.7 K and 24.7 K through the McMillian-Allen-Dynes formula[31], which are much larger than the value in graphene. Although, the experimental realization of $β_{12}$ has been faithfully confirmed by several research groups, there is still no experimental probe about the superconductivity of it for now. From first-principles calculations, Cheng *et al* argued that the expected high-Tc of $β_{12}$ is effectively suppressed by the substrate-induced strain and electron doping, can be enhanced by compressive strain and hole-doping[32].

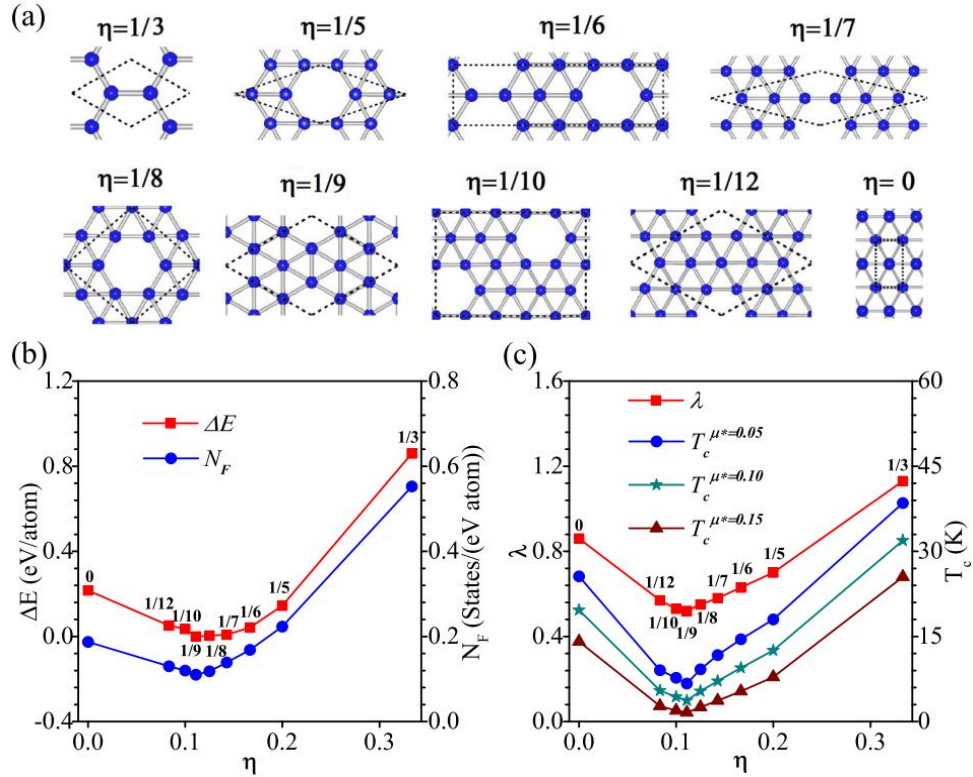

**Fig. 1.** (a) Most stable borophenes for each given hexagon hole density *η*. (b) Total energy and $N_F$ vs *η*. (c) *λ* and $T_c$ vs *η*，with permission from [13].

## 2.2 Topological properties

The discovery of 2D graphene and the Dirac fermion feature it possessed has constantly inspired people in the last decade to search for more candidates of 2D Dirac materials[33, 34]. Nowadays, 2D Dirac materials have spread to plenty materials, not bound to carbon materials only. Boron materials have the potential to spawn a great deal of 2D Dirac materials for its structural diversity and complexity as the left side element of carbon atom[35].

In 2014, Zhou *et al* proposed a 2D boron sheet named *Pmmn* boron with massless Dirac fermions using *ab initio* evolutionary structural search method, which is the first 2D Dirac boron material[15]. Since its electron-deficient feature, boron materials could present different bonding characteristics as compared with that of carbon materials. For instance, Ma *et al* reported a 2D partially ionic boron that consists of graphene-like plane and $B_2$ pairs that act as "anions" and "cations", respectively; In addition, this ionic structure exhibits double Dirac cones near the Fermi level with a high Fermi velocity that is even higher than that in graphene[17]. There are few other works about Dirac fermions properties in 2D boron, such as a series of planar boron allotropes with honeycomb topology[36], Dirac nodal-lines and titled Semi-Dirac cones coexisting in a striped boron sheet[37]. The one of the most inspiring findings of Dirac fermions in 2D boron allotropes is the successfully and firstly experimental identification of Dirac fermions in $\beta_{12}$ boron sheet growth on Ag(111)[38]; as presented in Figure 2, they argued that the lattice $\beta_{12}$ sheet can be decomposed into two triangular sublattices, analogues to the honeycomb lattice and, thus, host Dirac cones. Moreover, each Dirac cones can be split by introducing periodic perturbations. The current studies of topological properties have gone beyond the paradigm of Dirac fermions or Weyl fermions, been extended to multiple degenerated fermions in solid lattices, in which the Lorentz symmetry is not necessary to be respected[39]. Recently, Motohiko Ezawa presented there could be triplet fermions in $\beta_{12}$ boron sheet through altering its atomic interaction parameters and structural symmetry[40].

Since boron atom is light atom, the spin-orbit coupling (SOC) should negligible in boron materials, what makes their topological classifications are quite different from those of strong spin-orbit coupling materials, as discussed in the topological properties in carbon materials[25, 26]. As for superconductivity, the search for high-Tc materials has lasted for very long time, elemental boron materials or boron compounds may be one of the greatest potential candidates for their strong electron-phonon coupling. Considering the huge interests in topological properties and superconductivity in solid state materials, we have reasonable expectation that there should be more astonishing properties waiting for been discovered in boron materials.

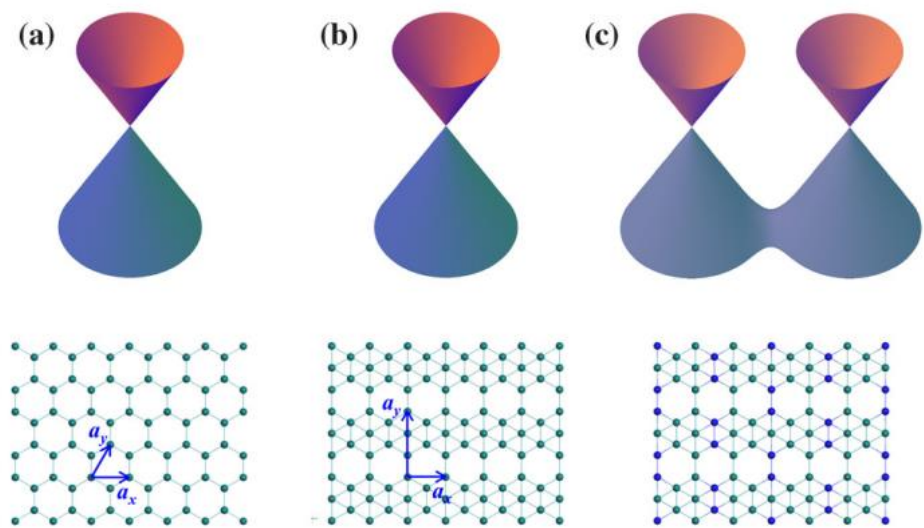

**Fig. 2.** The top and bottom panels are the band structures and corresponding atomic structures, respectively. (a) Honeycomb lattice. (b) $\beta_{12}$ sheet. (c) The $\beta_{12}$ sheet with a 3×1 perturbation, with permission from [38]

## 3 Thermal transport in borophene

Since the experimental realization of two-dimensional boron sheet (borophene), thermal transport in different phases of borophene becomes an attractive topic nowadays. The highly anisotropic thermal transport and moderate thermal conductivity make borophene promising applications in transparent conductor and thermal management. It was found that thermal transport among different phases of borophene are quite different, e.g. the thermal stability, thermal conductivity, anisotropy, etc. What's more, we also found different kinds of theoretical methods are adopted by these literatures, including phonon Boltzmann transport equation (pBTE), nonequilibrium Green's function (NEGF), and molecular dynamics (MD) simulations. Besides, some efforts were dedicated to the strain effect and mechanical properties of borophene. In this part, we will give a general view on the thermal transport in borophene by summarizing these achievements made in recent years, with an eye to understand the transport behavior in borophene and to call for further study on these materials.

### 3.1 Basic theories of thermal transport

Thermal transport is external phenomenon of lattice vibration, specifically, characterized by the phonon transport behaviors. Macroscopically, thermal transport means the magnitude of heat flux when a conductor is placed in a temperature gradient, while complex phonon transport and scattering mechanisms are taken place in this process microscopically. Different kinds of method had been developed to calculate the thermal transport properties of materials, including phonon Boltzmann transport equation (pBTE), molecular dynamics (MD) simulations, and nonequilibrium Green's function (NEGF), etc. These methods show distinct features and functions. For instance, the MD simulations can simulate the thermal transport in a system with large scale and size, which simulate more like real experiment. The pBTE is a century old method based on a primitive cell and the interatomic force constants (IFCs), being more applicable and extensive. We now give a general introduction of these methods. At first, we would like to introduce pBTE since it has been developed by Li Wu *et al* and widely used nowadays.[41-43] At thermal equilibrium, in the absence of external forces, the phonons are distributed according to Bose-Einstein statistics $f_0$. when there is a temperature gradient $\nabla T$, phonon distribution $f$ deviates from $f_0$, and this process can be described via BTE:

$$\nabla T \cdot \vec{v}_\lambda \frac{\partial f_\lambda}{\partial T} + \frac{f_\lambda - f_0}{\tau_\lambda^0} = 0, \quad (1)$$

the first term means the phonon diffusion due to $\nabla T$, the second term denotes the phonon scattering arising from allowed processes. The solution of this BTE is trivial because of the complex phonon scattering processes, i.e. the relaxation time $\tau_\lambda^0$ cannot be accurately determined. Considering two- and three-phonon processes as the only scattering sources, Li Wu *et al* defined $\vec{F}_\lambda = \tau_\lambda^0(\vec{v}_\lambda + \Delta_\lambda)$, $\tau_\lambda^0$ and $v_\lambda$ are the relaxation time and group velocity of phonon mode λ. If $\Delta_\lambda$ is set to zero, the equation is equivalent to working within the relaxation time approximation (RAT). Thus, $\Delta_\lambda$ is the measure of how much population of a specific phonon mode. $\tau_\lambda^0$ and $\Delta_\lambda$ can be computed as

$$\Delta_\lambda = \frac{1}{N} \sum_{\lambda'\lambda''}^{+} \Gamma_{\lambda\lambda'\lambda''}^{+} (\zeta_{\lambda\lambda''} F_{\lambda''} - \zeta_{\lambda\lambda'} F_{\lambda'}) \\ + \frac{1}{N} \sum_{\lambda'\lambda''}^{-} \frac{1}{2} \Gamma_{\lambda\lambda'\lambda''}^{-} (\zeta_{\lambda\lambda''} F_{\lambda''} + \zeta_{\lambda\lambda'} F_{\lambda'}) + \frac{1}{N} \sum_{\lambda'} \Gamma_{\lambda\lambda'} \zeta_{\lambda\lambda'} F_{\lambda'} \quad (2)$$

$$\frac{1}{\tau_\lambda^0} = \frac{1}{N} (\sum_{\lambda'\lambda''}^{+} \Gamma_{\lambda\lambda'\lambda''}^{+} + \sum_{\lambda'\lambda''}^{-} \frac{1}{2} \Gamma_{\lambda\lambda'\lambda''}^{-} + \sum_{\lambda'} \Gamma_{\lambda\lambda'}) \quad (3)$$

where $\Gamma_{\lambda\lambda'\lambda''}^{\pm}$ denote the three-phonon scattering rates, i.e. the absorption and emission processes of phonons. To obtain these scattering rates, we need the scattering matrix, which depends on the anharmonic IFCs. $\Gamma_{\lambda\lambda'}$ is the scattering probabilities from isotopic disorder. The boundary scattering is not included in this method since it is a low-temperature behavior. At last, the lattice thermal conductivity $\kappa_l$ can be obtained in term of $\vec{F}_\lambda$ as

$$\kappa_l^{\alpha\beta} = \frac{1}{k_B T^2 \Omega N} \sum_\lambda f_0(f_0+1)(\hbar\omega_\lambda)^2 v_\lambda^\alpha F_\lambda^\beta \quad (4)$$

where Ω is the volume of unit cell. This method is implemented in ShengBTE code,[43] which requires second- and third-order IFCs as input, and can simulate bulk and even low-dimensional system.

The MD simulations usually include equilibrium MD (EMD) and nonequilibrium MD (NEMD).[44-47] The NEMD also contains two approaches, i,e. calculating heat flux by controlling the temperature gradient or conversely calculating temperature gradient by controlling the heat flux. There are some basic concepts in MD simulations, including time step, boundary condition, statistical ensemble, empirical potential function, etc. Specifically, the NEMD simulations usually go through three steps. The first is the relaxation stage, which makes the system fully relaxed and reaches the equilibrium state under fixed temperature. The second is the transitional stage, transiting a period by placing the equilibrium system in a statistical ensemble. The final one is the nonequilibrium stage, turning the system from equilibrium to nonequilibrium state with a stable heat flux by applying temperature gradient. And then, one can obtain the statistics of the heat flux and temperature distribution. The lattice thermal conductivity can be calculated according to the Fourier's law

$$\kappa_{ph} = \frac{J}{A \cdot \nabla T} ,  \quad (5)$$

where $A$ is the cross sectional area of the system, and $\nabla T$ is the temperature gradient. $J$ is the heat flux, which is obtained by the slop of exchange energy $E_{ex}$ to the time step

$$J = \frac{E_{ex}}{A\Delta t} \quad (6)$$

As we know, the thermal conductivity of bulk is independent of sample size, while it is another thing for low-dimensional systems.[48-50] Hence, calculating thermal conductivity of low-dimensional systems needs to consider different sample size in a large scale, and the MD simulations rightly show advantages in this kind of calculations. However, the shortage of empirical potential function impedes a wide application of this method to various systems.

The NEGF method for phonon transport, developed by Yamamoto *et al*, is more suitable for dealing with nanostructures.[51] This method is based on a double electrode model. When the model is placed between a hot and cold heat baths, the thermal current can be written as

$$J_{th} = \int_0^\infty \frac{d\omega}{2\pi} \hbar\omega [f_L(\omega) - f_R(\omega)] T_{ph}(\omega) \quad (7)$$

where $f_{L(R)}(\omega)$ is the Bose-Einstein distribution of equilibrium phonons with energy $\hbar\omega$ in the left (right) electrode with temperature $T_L(T_R)$, $T_{ph}(\omega)$ is the phonon transport function. $T_{ph}(\omega)$ can be obtained through the Green's function of the scattering region, which is calculated via the self-consistent of Hamiltonian matrix of the model. After the Green's function of the scattering region is converged, $T_{ph}(\omega)$ can be calculated as

$$T_{ph}(\omega) = Tr[\Gamma_L(\omega) G_S^r(\omega) \Gamma_R(\omega) G_S^a(\omega)] \qquad (8)$$

where $G_S^{r/a}(\omega)$ is the retarded/advanced Green's function of the scattering region. Here, $\Gamma_{L/R}(\omega) = i\left[\sum_{L/R}^r(\omega) - \sum_{L/R}^a(\omega)\right]$, and $\sum_{L/R}^{r/a}(\omega)$ is the retarded/advanced self-energy due to the coupling to the left/right electrode. In the limit of small temperature difference between the hot and cold heat baths, the thermal conductance $\kappa_{ph} \equiv J_{th}/(T_L - T_R)$ is given by[51, 52]

$$\kappa_{ph}(T) = \frac{\hbar^2}{2\pi k_B T^2} \int_0^\infty d\omega \omega^2 T_{ph}(\omega) \frac{e^{\hbar\omega/k_B T}}{\left(e^{\hbar\omega/k_B T} - 1\right)^2} \qquad (9)$$

### 3.2 Thermal stability and conductivity of borophene

Among all the different types of boron sheets predicted theoretically,[19] the phases of $\beta_{12}$, $\chi_3$, and $\delta_6$ are experimentally reported in recent years, all grown epitaxially on Ag(111) substrate and showing weak interaction with the substrate.[22, 23] However, these boron sheets may be not thermally stable after leaving the substrate, especially for the $\delta_6$ phase. In spite of this reality, people have taken much effort to investigate the thermal transport in $\delta_6$ phase borophene. We summarized the main results as listed in Table 1. The $\delta_6$ phase borophene has two atoms per unit cell with the space group *Pmmn*, along armchair direction are parallel linear chains while a remarkable buckling is observed along the zigzag direction, indicating a high anisotropy. The experimental lattice constants of are a=5 Å and b=2.89 Å.[22] As compared to experimental borophene with the substrate, it can be found in Table 1 that the optimized parameter of free-standing borophene is almost unchanged along zigzag (b-axis) direction while only about 1/3 to the experiment along armchair direction (a-axis).

The phonon spectrum of borophene ($\delta_6$) obtained by finite displacement (FDS) method and density function perturbation theory (DFPT) are shown in Figure 3.[53, 54] The two boron atoms in the unit cell form six dispersion branches, including three acoustic and three optical branches, the same as graphene, silicene, phosphorene, etc.[55] In Figure 3, both FDS and DFPT point to an imaginary frequency of ZA branch along Γ-X (armchair) direction, which indicates the instability of the long-wavelength transverse thermal vibrations. This phenomenon in turn may explain the stripe formation along the armchair direction in the synthesis of borophene.[54] In contrast to previous two-dimensional (2D) materials, the phonon dispersion of borophene shows a highly anisotropy along armchair (Γ-X) and zigzag (Γ-Y) directions. Importantly, the optical branches along Γ-X are more dispersive than that along Γ-Y, indicating the high optical phonon velocities along the armchair direction, which would remarkably contribute to the thermal conductivity. Besides, Sun *et al* pointed to the stronger B-B bonding along armchair direction by analyzing the polarized vibrations.[54] All these characteristics are associated with the thermal transport behaviors.

In order to deal with the instability of $\delta_6$ borophene, i.e. the imaginary frequency near Γ point, two efficient empirical potentials for MD simulations were developed to describe the interaction of borophene with low energy triangular structure by Zhou *et al*.[57] One is the linear potential valence force filed (VFF) mode, while the other is nonlinear Stillinger-Weber (SW) potential. The VFF mode is useful for the description of interatomic interactions in covalent materials, in which the interactions are associated with some bond stretching and angle bending components. The SW potential is one of the most efficient nonlinear potentials, which includes both linear and nonlinear interactions. The linear component of the SW potential can be derived based on the VFF mode. As shown in Figure 3c, the imaginary frequency near Γ point disappears since the long-range interaction is not considered in these mode and potentials. Their MD simulations were performed within the publicly available package LAMMPS. Another strategy of removing the instability is full hydrogenation, and the hydrogenated product is now called borophane. After the full hydrogenation, there are two boron atoms and two hydrogen atoms in the unit cell, with an increased lattice constant *a* and an almost unchanged lattice constant *b* (Table 1), as compared to borophene.[58] The increased lattice constant *a* corresponds to the significantly stretched $B_1$-$B_1$ bond, which is 1.613Å in borophene while 1.935Å in borophane.[58] From Figure 3d, one can notice that all the phonon frequencies are positive in the whole Brillouin zone, showing that the hydrogenated borophene (borophane) is dynamically stable. Thus, the hydrogenation can effectively improve the stability of borophene. It is interesting that the hydrogen atoms dominate the high frequency branch and show weak interaction with lower phonon branches due to its light atomic mass.

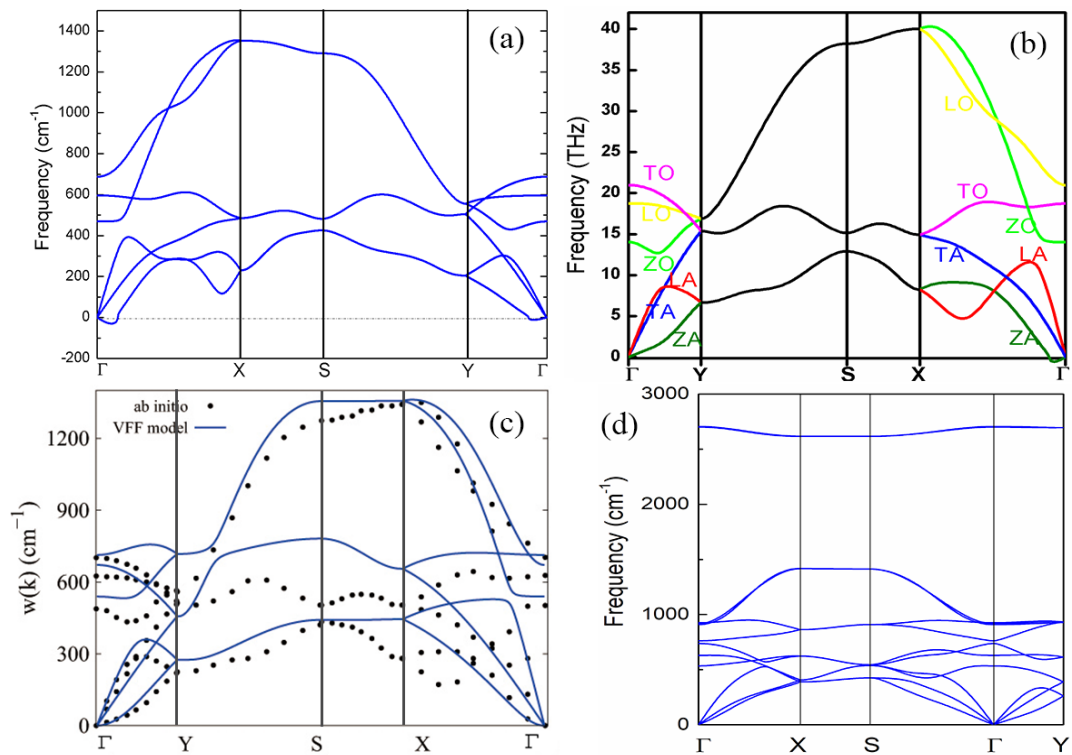

**Fig. 3.** Phonon spectrum of $\delta_6$ borophene. (a) and (b) are individually calculated by density function perturbation theory (DFPT) and finite displacement (FDS) method, adapted with permission form [53] and [54], respectively. And (c) is the phonon spectrum calculated by molecular dynamics simulations within valence force filed (VFF) mode, compared to the result of *ab-initio* calculation, with permission from [57]. (d) The phonon spectrum of fully

hydrogenated borophene (borophane), adapted from [58].

In addition to the $\delta_6$ phase borophene, people also concentrated on the thermodynamic properties of borophene with other phases, i.e. the $\beta_{12}$, $\chi_3$, $\alpha$ and $\alpha'$ phases.[56, 59, 60] To our surprise, these phases seems like more stable than $\delta_6$ since their phonon spectrum has no imaginary frequency. According to the different symmetry of the geometric structure, their transport behaviors are also different. For example, the $\alpha$ and $\alpha'$ phases are isotropic with a symmetrical phonon dispersion, while the anisotropy of $\delta_6$, $\beta_{12}$ and $\chi_3$ phases is observed through the difference in the phonon dispersion around $\Gamma$ between [$\Gamma$-X] and [$\Gamma$-Y]. As can be found in previous calculations, the allowed phonon frequencies for different phases of borophene are quite similar since the components are the same B atoms. However, different number of B atoms per unit cell results in different number of optical phonon branches, along with the distinct geometric symmetry, lead to multiple thermal transport in these borophene.[19, 59] Thermal conductivity is a directly index to describe the phonon transport properties, which provides some guidance for possible applications of the materials. For instance, a high thermal conductivity is necessary to removing the accumulated heat in electronic or photovoltaic devices, while low thermal conductivity is pursued by thermoelectric and thermal insulation community.

We also collected the available thermal conductivity values of borophene at room temperature calculated by different kinds of methods, as shown in Table 1. It was found that the thermal conductivity of borophene is much lower than that of graphene. The thermal conductivity of $\delta_6$ borophene at room temperature is 43.3 ($10^{-9}$ W/K) and 22.6 ($10^{-9}$ W/K) along armchair and zigzag directions,[54] respectively, calculated through pBTE, which are quite lower than 1775 ($10^{-9}$ W/K) of graphene.[61] The low thermal conductivity of borophene arises mainly from the small extension of acoustic dispersion with the low group velocities. One notices that the thermal conductivity along zigzag direction is about half of that along armchair direction, the higher value in armchair direction can be attributed to the stronger B-B bonding and the highly dispersive optical branches along this direction.[54] Zhou *et al*[56] calculated the thermal conductance of borophene by ballistic NEGF method. They calculated the ballistic phonon transport and found superior lattice thermal conductance of borophene, which helps to understand the size effect and transport behavior in borophene. In contrast to borophene, borophane possesses a higher thermal conductivity along zigzag direction, which means that the full hydrogenation of borophene change the anisotropy of the thermal transport, this unusual phenomenon are confirmed by group velocity and scattering mechanism, as presented by Liu *et al*.[58] It can also see in Table 1 that the thermal conductivity calculated by different method are much similar, e.g. the results of $\delta_6$ obtained by pBTE and MD are very close to each other.[54, 62] We also found that the α′ borophene has an isotropic and low thermal conductivity of 5.96 ($10^{-9}$ W/K),[60] which behaves more like that in monolayer TMDCs, e.g. monolayer $ZrS_2$, $HfS_2$, etc.[63, 64] The low thermal conductivity of α′ compared to $\delta_6$ is due to a large number of optical branches and also the strong coupling of acoustic and optical branches with high

| Phase | a (Å) | b (Å) | $\kappa_l$ (a, b) ($10^{-9}$ W/K) | Thermal stability | Method |
|---|---|---|---|---|---|

scattering rates. The diversity of thermal transport in different borophene makes it possible applications in transparent conductor and thermal management.

**Table 1.** The lattice parameters, room temperature thermal conductivity, thermal stability and computational method of borophene from related literatures. Some thermal conductivity values in corresponding literature are unavailable. Not that the unit of thermal conductivity is normalized in 2D form.

| | | | | | |
|---|---|---|---|---|---|
| | [22]5 | 2.89(exp.) | | | |
| | [53]1.613 | 2.864 | | No | |
| | [54]1.617 | 2.872 | (43.3, 22.6) | No | pBTE |
| $\delta_6$ | [56]1.612 | 2.869 | | No | NEGF |
| | [62]2.1 | 3.18 | (42.6, 22.01) | No | MD |
| (VFF/SW) | [57]1.614 | 2.866 | | Yes | MD |
| (Hydrogen) | [58]1.93 | 2.82 | (79.7, 147.8) | Yes | pBTE |
| $\alpha'$ | [60]4.37 | 4.37 | 5.96 | Yes | pBTE |
| $\beta_{12}$ | [56]5.062 | 2.924 | | Yes | NEGF |
| 8-*Pmmn* | [65]4.52 | 3.25 | (61.9, 86.1) | Yes | pBTE |

From discussions above, we know that thermal transport in different borophene exhibits special properties. Specifically, some of which are highly anisotropic while other are isotropic; some are intrinsically stable while some are thermally unstable; the magnitude of thermal conductivity shows much difference owing to the different phonon scattering rates. It was noticed that studies on the thermal transport of borophene are still lacking since the thermal conductivity of some phases of borophene are unknown, which call for further studies. We now take a look at the size dependency of the thermal conductivity. The phonon mean free path of $\delta_6$ borophene, calculated via pBTE, are 300 nm and 400 nm for the armchair and zigzag directions, respectively,[54] which are six orders of magnitude smaller than that of graphene. The short mean free path of borophene also point to its lower thermal conductivity. In MD simulations, calculated thermal conductivity usually show a strong size dependency until the sample length is smaller than the allowed phonon mean free path.[62] The thermal conductivity by MD increase with increasing sample length, and the converged thermal conductivity is close to that by pBTE. At last, it is worthwhile to note that 8-*Pmmn* borophene was taken as an example to confirm the intrinsic quadratic acoustic branch of 2D systems and also its effect on thermal conductivity.[65] Using physical IFCs, they shown that a quadratic dispersion phonon branch is always presented on suspended few-layer systems, which can have a strong impact on the thermal conductivity of 2D systems. The physical IFCs result in a negligible change of the phonon dispersion, except for turning the lowest acoustic mode to a quadratic dispersion near Γ, which has a dramatic effect on the thermal conductivity, decreasing by 50%. Hence, the quadratic dispersion of the lowest phonon branch is a universal feature of 2D systems.

### 3.3 Strain effect on borophene

The instability of borophene ($\delta_6$) made people try to study its mechanical properties by applying tensile strains. Both Pang and Wang have calculated the strain effect on the mechanical flexibility and phonon instability.[66, 67] Their results are much similar, both indicated that the mechanical properties along armchair and zigzag direction are highly anisotropic. As shown in Table 2, when applying uniaxial strains, the critical strain along armchair and zigzag directions are 0.08 and 0.15, respectively, showing the superior flexibility along zigzag direction, which corresponds to the higher flattened strain (by 27%) of the bucking height along this direction. As for the biaxial strain, the critical strain is 0.08.[67] For both uniaxial strain along armchair and biaxial strain, the phonon spectrums are dynamically unstable when the bond lengths are extended by 8%, suggesting a lower limit for which the structure remains stable along the out-of-plane direction. In contrast, such phonon mode remains stable under uniaxial strain along zigzag direction. No matter what kind and how strong the strain been applied, the imaginary frequency cannot be fully removed, which indicates that the freestanding borophene is mechanically unstable even with strains, and the stability of synthesized borophene is probably provided by the out-of-plane constraints from the silver substrates.

In Table 2, borophene can withstand stress up to 20.26 N/m and 12.98 N/m in armchair and zigzag directions, respectively. The phonon calculations indicate that under uniaxial tension along the zigzag direction, the failure is attributed to the elastic instability, whereas along armchair direction and biaxial tension the failure mechanism is phonon instability and such instability is dictated by the out-of-plane acoustic mode.[67] As for comparison, the critical strengths of borophene are obviously higher than that of silicene and black phosphorus,[68, 69] while much smaller than that of graphene,[70] which because of the fact that B-B bonds are stronger than Si-Si and P-P bonds, while weaker than C-C bonds in graphene. It was found that the critical strain of borophene may be the lowest among all the studied 2D materials. The failure mechanisms of silicene and graphene are similar, that is the elastic instability under both uniaxial and biaxial tensions. While the failure mechanisms of borophene are found to be similar to $MoS_2$,[71] with one of the elastic instability while the rest two of the phonon instability. What's more, the critical strengths and strains of borophene are more anisotropic than that of silicene and graphene because of their hexagonal structures.

As we know, borophene can be stabilized by hydrogenation, with the product called borophane. Wang *et al* have calculated the mechanical properties and phonon stability of strained borophane.[72] They considered uniaxial tensile strains along armchair and zigzag directions, and also the biaxial strains. The critical strains are 0.12, 0.3, and 0.25 for uniaxial strain along armchair and zigzag directions, and biaxial strain, respectively, see in Table 2. Thus, the mechanical properties of borophane are also highly anisotropic, and armchair direction also possesses the superior mechanical flexibility. Calculated phonon dispersions under different strains indicate that borophane can withstand up to 5%, 15% uniaxial tensile strains along armchair and zigzag directions, respectively, and 9% biaxial strains, which means that borophane may become unstable before reaching the critical strains.[72] As one notices, the strain that borophane can withstand along the zigzag direction is three times of that along the armchair direction. Consequently, borophane exhibits better mechanical stability and phonon stability along the zigzag direction.

**Table 2.** Lattice parameters, critical tensile strain and corresponding ultimate strength for different borophene. The strains been applied include uniaxial strain along armchair direction, uniaxial strain along zigzag direction, and the biaxial strain.

|  | a (Å) | b (Å) | Buckling height (Å) | Uniaxial (armchair) | Uniaxial (zigzag) | Biaxial |
|---|---|---|---|---|---|---|
|  |  |  |  | Critical tensile strain | | |
|  |  |  |  | Ultimate strength (N/m) | | |
| $\delta_6$ | [67]1.614 | 2.866 | 0.911 | 0.08 | 0.15 | 0.08 |
|  |  |  |  | 20.26 | 12.98 | 14.75 |
| Borophane | [72]1.941 | 2.815 | 0.81 | 0.12 | 0.3 | 0.25 |
|  |  |  |  | 12.26 | 18.48 | 21.06 |
| $\beta_{12}$ | [74]5.062 | 2.924 | 0 | 0.2 | 0.21 |  |
|  |  |  |  | 19.97 | 20.38 |  |
| $\chi_3$ | [74]2.9 | 4.44 | 0 | 0.21 | 0.155 |  |
|  |  |  |  | 19.91 | 20.18 |  |
| $\alpha$ | [74]5.046 | 5.044 | 0.17 | 0.21 | 0.16 |  |
|  |  |  |  | 14.84 | 18.83 |  |

Recently, Mortazavi *et al.* have studied the strain effect on the thermal conductivity of borophene by MD simulations.[62] The thermal conductivity at room temperature they calculated are 42.63 ($10^{-9}$ W/K) and 22.01 ($10^{-9}$ W/K) for armchair and zigzag directions, respectively, being consistent with the results of pBTE (Table 1). As for the strain effect, they observed that there is a significantly increase in thermal conductivity along armchair direction when the strain is applied in this direction, while there is just a smaller increase for zigzag direction.[62] However, they did not analyze in depth why it behaves like this. This different response is associated with the anisotropic crystal of borophene, which reinforces the prospective application in the construction of phononic device. It is worthwhile to note that borophene may be not suitable for thermoelectric applications since it has a metallic behavior and the thermal conductivity doesn't reach the required lower limit.

In Table 2, we also shown the stain effect and mechanical properties of other borophene members, i.e. $\beta_{12}$, $\chi_3$ and $\alpha$.[73, 74] Along armchair direction, their critical tensile strains are higher than that of $\delta_6$, indicating a better flexibility of these sheets along armchair direction. This also points to the possible phonon instability of $\delta_6$ along armchair direction against long-wavelength transversal waves, and these borophene members are more stable than $\delta_6$ borophene. While along zigzag direction, there critical strains are close to $\delta_6$ but still a little higher, with the higher ultimate tensile strengths along this direction. In contrast to $\delta_6$, for which the zigzag direction has superior flexibility, these boron sheets have better flexibility along armchair direction especially for $\chi_3$ and $\alpha$ sheets. This difference is attributed to the different atomic configurations and the way they evolve and rearrange during the loading condition.[74] Along armchair direction the ultimate strengths of $\beta_{12}$ and $\chi_3$ are very close to each other, of which the fully occupied hexagonal or zigzag lattices are connected by single B-B bonds. When stretched along zigzag direction, the fully occupied hexagonal lattices in $\beta_{12}$ can extend more than zigzag lattices in $\chi_3$ borophene.[74] These results can help to understand the stability and mechanical properties of borophenes.

### 3.4 Borophene nanoribbon

The discovery of borophene also draws immediate attention to the investigation of borophene nanoribbons (NBs). The magnetism, electronic and phonon transport properties of borophene NBs are predicted by first-principles calculations. It was found that pristine $\delta_6$ armchair NBs are non-magnetic, while zigzag NBs adopt a magnetic ground states, either anti-ferromagnetic or ferromagnetic depending on the ribbon width.[75] Upon hydrogenation, all turn to non-magnetic. Liu *et al* discovered negative differential resistance and magnetoresistance in zigzag borophene NBs.[76] Hence, borophene NBs can be built with different structures and magnetic moment. These electronic properties can be taken into account in the further for the design of devices involving these NBs.

Recently, several kinds of borophene NBs was synthesized on Ag (100) surface, which may promote further applications of borophene.[77] Jia *et al* have studied the thermal transport in borophene ($\delta_6$) NB by MD simulations.[78] The length of the sample they chosen for simulation is 40 nm, while the widths vary from 10 nm to 30 nm. The thermal conductivity of zigzag NBs is much higher than armchair NBs, i. e. 586.12 W/mK and 257.62 W/mK, respectively. It was found that the thermal conductivity of the NBs shows weak dependence on the widths, and also shows little sensitivity to the strains. The thermal conductivity doesn't change as the strains ranging from 0.5% to 2%, applying along the heat flux direction. Zhang *et al.* also studied the phonon transport in borophene NBs with other phases, i. e. $\beta_{12}$ and $\chi_3$, both exhibiting diverse phonon transport abilities and different anisotropies.[79] They mainly compared the transport properties with that of graphene's, among theses NBs the highest thermal conductance is comparable to that of graphene's, whereas the lowest thermal conductance is less than half of graphene's. Overall, the experimental synthesis and the novel thermal properties of borophene NBs enrich the low-dimensional allotrope of boron and possible applications in thermal management devices.

As a closing of this part, we try to highlight the special thermal transport in different phases of borophene. A general view of the thermodynamic, mechanical properties and even the borophene nanoribbon are given, which may provide a reference for further studies on the thermal transport of borophene. However, to our knowledge, the instability of borophene is still not fully understood, and also the anisotropic response of the thermal transport to the strain effect is not clear. What's more, the achievements made in recent years are mainly based on the striped $\delta_6$ borophene, thermal transport in some of the boron sheets are still unknown. These questions call for continuous study on the thermal transport and mechanical properties of borophene.

## 4 Conclusion

As we can see from this review, the investigation of borophene is growing explosively. At first, we mainly focus on the crucial electronic properties of borophene, i.e., the superconductivity and topological properties. The strong electron-phonon coupling and metallic nature of borophene make it promising candidate of superconductivity. Meanwhile, borophene is also observed as two-dimensional Dirac material with high Fermi velocity. Following the electronic properties, we also refer to thermal transport properties of borophene, a comparison of the thermal transport and mechanism properties of different borophene has been made, displaying the potential application of borophene on transparent conductor and thermal management. Borophene exhibits various investigations and applications of certain community. In this work, we merely give a general view on the thermal and electronic properties, which point to its promising applications in thermal conductor, nanoelectronic devices, etc. However, most of these achievements were made theoretically through first-principles calculations. Further experiments should focus more on these novel physical properties.